\documentclass[journal=nalefd,manuscript=letter]{achemso}
\usepackage[english]{babel}
\usepackage{graphicx}
\usepackage{amsmath}
\usepackage{color}
\graphicspath{{Fig/}}

\title{ Effect of nematic ordering on electronic structure of FeSe}

\author{A.~Fedorov}

\affiliation{IFW Dresden, P.O. Box 270116, D-01171 Dresden, Germany}
\alsoaffiliation{II. Physikalisches Institut, Universit\"at zu K\"oln, Z\"ulpicher Strasse 77, 50937 K\"oln, Germany}
\alsoaffiliation{St. Petersburg State University, St. Petersburg, 198504, Russia}
\author{A. Yaresko}
\affiliation{Max Planck Institute for Solid State Research, Heisenbergstrasse 1, 70569 Stuttgart, Germany}

\author{T. K. Kim}
\affiliation{Diamond Light Source, Harwell Campus, Didcot OX11 0DE, United Kingdom}

\author{E. Kushnirenko}
\affiliation{IFW Dresden, P.O. Box 270116, D-01171 Dresden, Germany}

\author{E. Haubold}
\affiliation{IFW Dresden, P.O. Box 270116, D-01171 Dresden, Germany}

\author{T. Wolf}
\affiliation{3Institute for Solid State Physics, Karlsruhe Institute of Technology, 76131 Karlsruhe, Germany}

\author{M. Hoesch}
\affiliation{Diamond Light Source, Harwell Campus, Didcot OX11 0DE, United Kingdom}

\author{A.~Gr\"uneis}
\affiliation{II. Physikalisches Institut, Universit\"at zu K\"oln, Z\"ulpicher Strasse 77, 50937 K\"oln, Germany} 

\author{B.~B\"uchner}
\affiliation{IFW Dresden, P.O. Box 270116, D-01171 Dresden, Germany}

\author{S. V. Borisenko}
\affiliation{IFW Dresden, P.O. Box 270116, D-01171 Dresden, Germany}

\email{s.borisenko@ifw-dresden.de}

\begin{document}
\clearpage
\begin{abstract}
Electronically driven nematic order is often considered as an essential ingredient of high-temperature superconductivity~\cite{Fernandes2014,Fernandes2013,Livanas2015,Yang2013}. Its elusive nature in iron-based superconductors resulted in a controversy not only as regards its origin~\cite{Fernandes2014,Baek2014,Zhang2015}  but also as to the degree of its influence on the electronic structure even in the simplest representative material FeSe~\cite{Maletz2014,Shimojima2014,Nakayama2014,Zhang2015,Borisenko2015,Watson2015a,Fanfarillo2016,Xu2016,Ye2015} . Here we utilized angle-resolved photoemission spectroscopy and density functional theory calculations to study the influence of the nematic order on the electronic structure of FeSe and determine its exact energy and momentum scales. Our results strongly suggest that the nematicity in FeSe is electronically driven, we resolve the recent controversy and provide the necessary quantitative experimental basis for a successful theory of superconductivity in iron-based materials which takes into account both, spin-orbit interaction and electronic nematicity.
\end{abstract}

%\maketitle

\section{Introduction}

Identification of the driving force behind an ordering phenomenon in a solid is one of the most fundamental questions of the condensed matter physics. Successful theory of such a phase transition to the ordered state should be able to predict the magnitude of the critical parameter on the basis of the characteristics of the normal state, i.e. when the ordering has not yet occurred, and single out the interaction which triggers the transition. In the ordered state many degrees of freedom, such as lattice, charge, spins or orbitals will respond and identification of the culprit becomes more difficult.  The solution of this problem for the nematic phase of iron-based superconductors (IBS) has a special meaning since it may shed light on the origin of the electronic pairing itself. 
FeSe is a very convenient system for the identification of the interaction governing the tetragonal-to-orthorhombic transiton since static magnetic ordering, which breaks time-reversal symmetry in other IBS at lower temperatures, is not observed. The debates, however, are still very lively even concerning this material. First, it is not entirely clear whether this nematic transition is of electronic or lattice origin. Second, if it is the former, is it spin- or orbital driven? Knowing the answer to the last question would assist one to choose between sign-preserving or sign-changing superconducting order parameter.

The direct way to understand the nature of the transition is to study its influence on the electronic structure of FeSe. Recent ARPES results are controversial: from one side~\cite{Shimojima2014, Nakayama2014,Zhang2015,Watson2015a,Fanfarillo2016,Xu2016,Ye2015}  a gigantic energy splitting of the order of 50 meV ($\approx$600K) has been reported to occur below the structural transition at Ts = 87 K; from the other side, no influence of nematicity on the electronic structure has been directly resolved~\cite{Maletz2014}, at least on the scale larger than ~10 meV~\cite{Borisenko2015}.
In this study we explore the electronic structure of FeSe by using high-resolution ARPES and first principle DFT calculations with a new precision. These methods provide an explicit picture of the FeSe electronic structure and its evolution through the fourfold symmetry breaking phase transition in the broad temperature range. Our results suggest that the experimental electronic structure of FeSe cannot be fully explained by usual lattice deformation thus clearly implying the electronic origin of the nematic transition. Furthermore, we precisely determine the energy and momentum scales of the variations of the spectral function below Ts which remarkably agree with this temperature scale (87 K) and speculate about their magnetic or orbital origin.

\section{Breakdown of tetragonal symmetry.}

\begin{figure}[h!]
	\centering
		\includegraphics[width=16cm]{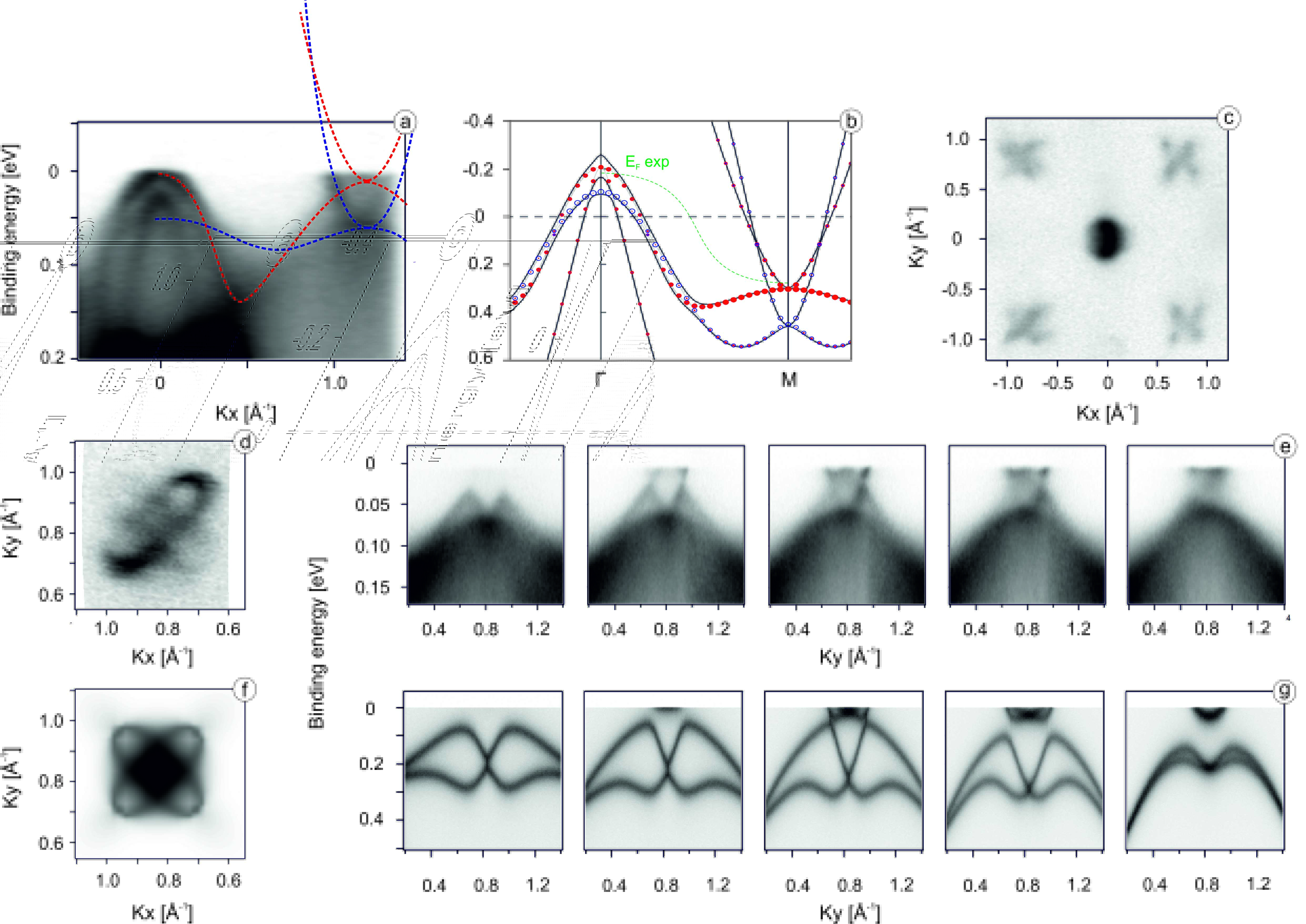}
	\caption{ARPES data compared to corresponding band-structure calculations of the tetragonal phase of FeSe: a) ARPES intesity along the diagonal of the BZ; b) full relativistic (solid lines) and scalar-relativistic (symbols) calculations for the tetragonal phase; red symbols represent $d_{xz,yz}$ states, blue symbols represent $d_{xy}$ states;  c) ARPES-derived Fermi surface map acquired at $h\nu$=100eV showing the full Brillouin zone with the hole-like pocket in its center and electron-like pockets in its corners; d) high-resolution Fermi surface map of electron pockets measured at $h\nu$=28eV; e) momentum-energy intensity maps taken at 0.2  ,0.15, 0.1, 0.5, 0 \AA$^{-1}$ from the center (0.82\AA$^{-1}$) of electron pockets; f) electron pockets simulated by the integration within 5 meV of band structure at 250 meV binding energy of bare DFT results;  e) corresponding to panel d calculated momentum-energy intensity maps.}
	\label{Fig:elstruct}
\end{figure}

 To understand the general features of the electronic structure of FeSe in the ordered phase let us start by presenting a comparison of low temperature ARPES data and band structure calculations in Fig.~\ref{Fig:elstruct} a,b on a relatively large energy scale of hundreds of meV. Even at a first glance there is a good correspondence between the calculated in tetragonal phase (above transition) dispersions and experimental features. This qualitative agreement becomes quantitative if one performs two typical for IBS transformations: orbital-dependent renormalizations and relative energy shift of the constructions in the center  and the corner  of the Brillouin Zone (BZ). While the former transformation is due to the Hund-rule coupling and is well captured by the DMFT calculations~\cite{Kotliar2006,Valenti2012}, the latter is not yet well understood, but can be related either to the so called $s\pm$ Pomeranchuk instability~\cite{Chubukov2016,Yang2013} or particle$-$hole asymmetry~\cite{BenfattoPhysRevLett.103.046404}.  We note that such a shift, unlike the usually exercised “rigid” shift of the Fermi level, preserves the number of charge carriers in the system and results in singular Fermi surfaces in all known IBS \textit{e.g.}~\cite{Charnukha2015, Borisenko2015}. Position of the Fermi level corresponding to the experimentally observed one is schematically indicated in Fig.~\ref{Fig:elstruct} b by the dashed green line. Another effect responsible for the discrepancy between the experiment and calculations is the stronger renormalization of the $d_{xy}$ band. Obviously, if all the bands were renormalized by the same factor, there would be no qualitative difference between two panels, provided the above mentioned shift is absent. Comparing the states marked by blue color ($d_{xy}$) one notices reported before~\cite{Maletz2014} renormalization factor of 9, vs. “usual” factor $\approx$3 for the rest of the bands. The main observation here is that the electronic structure of FeSe measured deep in the ordered state (T $<$ Ts) is very reasonably described by the calculations in the tetragonal phase, meaning that the nematic order does not alter the electronic structure significantly.

The Fermi surface map of FeSe featuring the typical for IBS hole- and electron-like pockets is shown in Fig.~\ref{Fig:elstruct}c.
We further focus our attention on the electron pockets in the corners of the BZ and discuss the data taken along many different cuts in momentum, not only high-symmetry one, as in Fig.~\ref{Fig:elstruct} panel a. In Fig.~\ref{Fig:elstruct} d we compare the experimental Fermi surface map in the vicinity of the corner of the BZ with the calculated one (Fig.~\ref{Fig:elstruct} f ), simulated by the integration within 5 meV of slightly broadened in energy and momentum  (to mimic experimental resolution) band structure  at 250 meV binding energy of bare DFT results. Corresponding energy-momentum cuts taken at different values of $k_x$ are shown in Fig.~\ref{Fig:elstruct} e. Again, we observe a remarkable qualitative correspondence both in energy and momentum for all cuts. We note that the larger electron pocket appears much shallower in the experiment because of the stronger renormalization of the $d_{xy}$ band mentioned earlier. It is also clear now that the apparently elongated ellipses can be elongated already in the tetragonal phase, just because the experimental Fermi level runs close to the bottoms of the electron pockets and not because of the nematicity as was reported earlier~\cite{Shimojima2014, Nakayama2014,Zhang2015,Watson2015a,Fanfarillo2016,Xu2016,Ye2015}

The next step would be to compare the experiment with more exact calculations. Since the presence of the spin-orbit splitting has been demonstrated in all IBS~\cite{Borisenko2015} and crystal structure of FeSe below Ts is orthorhombic, we present the results of the fully relativistic band structure calculations with unequal in-plane lattice constants  (a=5.33426, b=5.30933)~\cite{Khasanov2010} in Fig.2 b. Inclusion of the spin-orbit interaction opens the gaps in the places of crossings of the non-relativistic dispersions, as has been reported previously~\cite{Borisenko2015,Maletz2014}. One example of such hybridization is seen in raw data in Fig.1a at approximately -0.25 \AA$^{-1}$, where the $d_{xy}$ states interact with $d_{xz}$, $d_{yz}$ states (intentionally not covered by schematic dashed lines).  In order to facilitate the further comparison with the experiment, we overlay the results for two perpendicular directions in the BZ (green and blue lines) in Fig. 2b. In this way we simulate the presence of two domains~\cite{Shimojima2014, Nakayama2014,Zhang2015,Watson2015a,Watson2015b} in the area on the surface of the sample covered by a typical source of photons in ARPES experiment. The results of the calculations in tetragonal phase (red lines) are shown as well, to illustrate what to expect from the structural transition. Although previously always referred to as “negligible” due to 0.4\% variation of the lattice constants , we clearly see the noticeable changes of the electronic structure both, when going from domain to domain and, what is more surprising, from tetragonal to orthorombically distorted  lattice. The most dramatic changes occur right at the corner of the BZ: some features, \textit{e.g.} the bottom of the deeper electron pocket at $\approx$0.4 eV calculated binding energy, split by 25 meV. Taking into account that the experimental Fermi level lies in this region, one should be able to observe such changes as a function of temperature, even in the presence of the band renormalization.

\begin{figure}[h!]
	\centering
		\includegraphics[width=17cm]{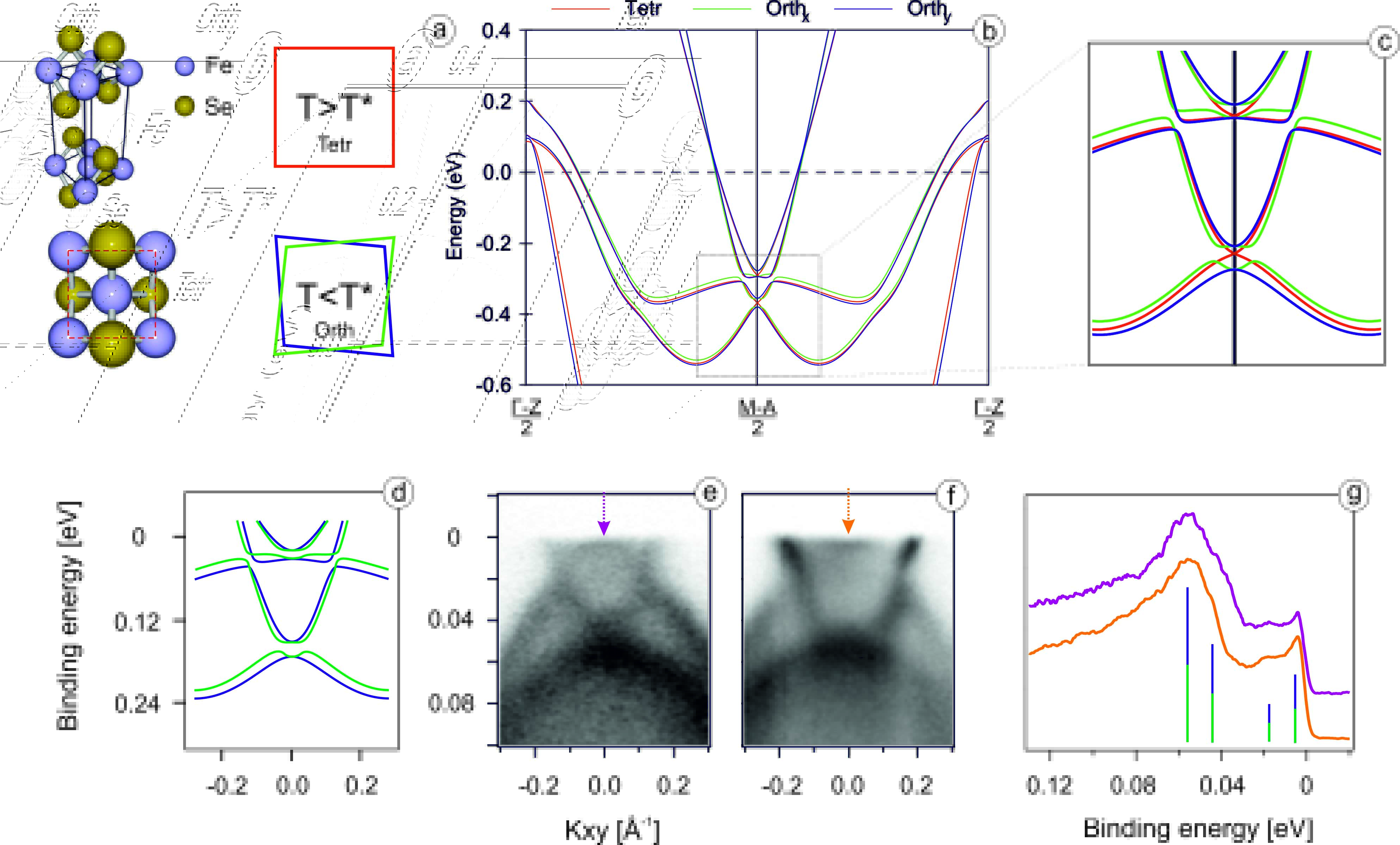}
	\caption{a) FeSe crystal structure; b) full relativistic DFT calculations for the tetragonal and orthorhombic phases of FeSe c) zoomed-in view of the DFT calculations close to the BZ corner; d) same as in panel c, but without tetragonal phase to simulate photoemission signal; e) and f) ARPES data along the $\Gamma - M$ direction in vicinity of the BZ corner taken at 6.4 K and using 42 eV and 28 eV photon energy, respectively; g) EDCs at the BZ corner obtained by integration within 0.06 \AA$^{-1}$ of ARPES data at panel e) and f) correspondingly.   }
	\label{Fig:TetrToOtrh}
\end{figure}

In Fig. 2~d and e,f we compare the zoomed-in region of the theory dispersions with the corresponding experimental data. We emphasize, that any kind of misalignment is ruled out since the data are taken from the maps of electron pockets recorded at different photon energies. Raw data clearly indicate the presence of the features, not seen in the calculations of the tetragonal phase. In particular, and this is directly seen in the energy distribution curves (EDC) shown in Fig. 2~g, each of the two bottoms of the electron pockets is indeed split, exactly as calculations predict. The value of this splitting is lower ($\approx$10 meV) than the theoretical one, as is expected because of the renormalization. Moreover, the fine structure expected from overlapping of two orthorhombic domains exactly corresponds to the experiment. The doubling of the features at higher binding energies is more difficult to see, but it is present in the second derivative plots (not shown). The obtained result complements our recent studies of the spin-orbit interaction in FeSe~\cite{Borisenko2015}. Earlier we have observed two relatively broad features in the EDC at the corner of the BZ and their splitting has not been directly resolved, leading us to an estimate of the upper limit of possible nematic splitting. Now we found the conditions at which we resolve the components directly and they are in a remarkable agreement with the band structure calculations taking into account both, spin-orbit interaction and orthorhombic distortion.

%Fig.~\ref{Fig:calc} shows the calculated dispersion among the $\Gamma - X$ direction, which consist three hole-like bands at the center and the electron-like bands in the corner of the Brillouin zone. 

\section{Temperature dependence of the electronic structure.}

Now we would like to trace the detected changes across the structural transition. First we provide the evidence that the mentioned above elongation of the electron pockets is indeed present in the normal state. The Fermi surface maps taken below (10 K) and well above (270K) the structural transition in Fig. 3 b,c both feature the electron pocket of elliptical shape. This means that the effects of nematicity are more subtle than it was believed earlier~\cite{Shimojima2014, Nakayama2014,Zhang2015,Watson2015a,Fanfarillo2016,Xu2016,Ye2015}. In Fig. 3d-f we present the much discussed temperature dependence of the EDC from the corner of the BZ. As we have demonstrated in Fig. 2 this EDC at low temperatures consists of four peaks, which can be made more visible by selecting particular experimental conditions (photon energy and geometry) to minimize the influence of matrix element effects. This is in contrast to what has been reported before~\cite{Shimojima2014, Nakayama2014,Zhang2015,Watson2015a,Fanfarillo2016,Xu2016,Ye2015}. Moreover, the EDC at higher temperatures has two components, as expected from the calculations of the tetragonal phase. Because of relatively high temperatures this splitting is not seen directly, but the characteristic lineshape of the EDC and its second derivative (Fig. 3f) clearly show the presence of two components. We note that the appearance of the especially prominent peak closest to the Fermi level is due to the two factors. First is the lower temperature which is known to expose sharp peaks residing close to the Fermi level because of the Fermi function which enters the expression for ARPES photocurrent. Second is the slightly asymmetric splitting of the bottom of the smaller electron pocket seen in the calculations of the orthorhombic phase (Fig. 2b). The binding energy of the closest to $E_F$ split component is only a few meV. We note that the splitting due to $a\neq b$ (~10 meV) is not very large and was therefore elusive before. It is not equally clearly seen at arbitrary experimental conditions (compare EDCs from Fig. 2 and 3). Sometimes it appears only as a shoulder near the more intense counterpart which is due to the matrix element effects. By now we can state that the effect of nematic ordering is indeed observed by ARPES, but its magnitude is of the order of 10 meV and what is more important, it is nicely described by the conventional band structure calculations which take into account spin-orbit interaction and orthorhombic distortion of the lattice. No attempts to simulate magnetism or orbital ordering have been made in the calculations. At this stage, one is tempted to conclude that the nematic phase in FeSe is just a consequence of the phonon-driven structural transition.

\begin{figure}[h!]
	\centering
		\includegraphics[width=17cm]{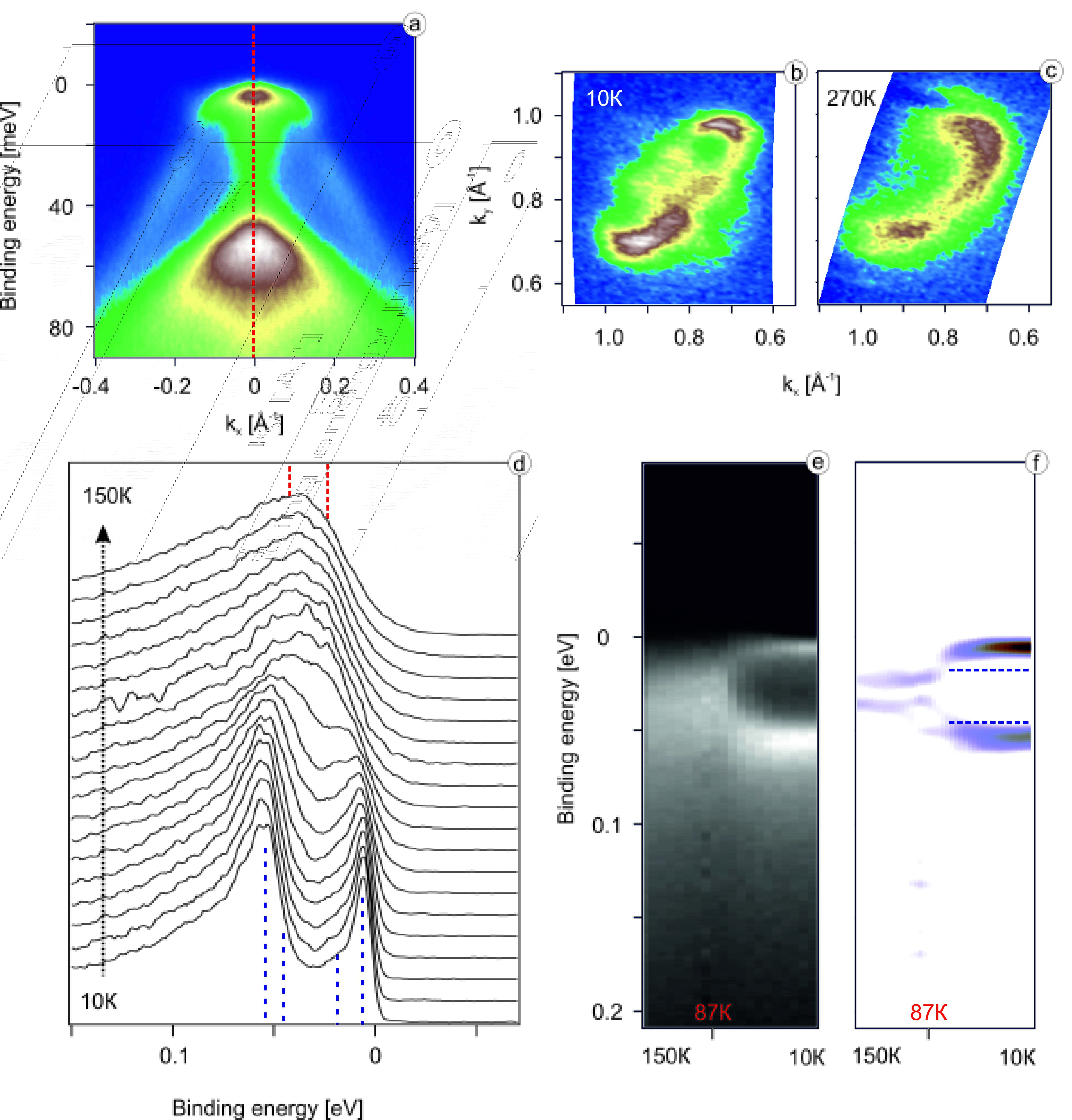}
	\caption{a) High resolution ARPES data along the shortest $A - A$ direction in vicinity of the $A$ point; b) high resolution ARPES derived Fermi surface in vicinity of $A$ point measuread with $h\nu$=28eV and T=10K; c) ARPES derived Fermi surface in vicinity of $A$ point measured with $h\nu$=28eV and T=270K; d) EDC curves taken at the $A$ point measured at temperatures from 10K to 150K; e) and f) color plot representation of data from panel d) and its second derivative, respectively. }
	\label{Fig:ElP}
\end{figure}

However, in order to draw the final conclusions we first have to investigate the center of the BZ with the same precision. Coming back to the results of the calculations in Fig.~2b, one easily notices that the red, blue and green dispersions forming the hole-like Fermi surfaces nearly coincide, in sharp contrast to the situation in the corner of BZ. This means that the calculations predict negligible variations of the electronic structure in this part of the reciprocal space both upon entering the tetragonal phase and between domains. In Fig.~4 we present corresponding experimental data. The high-precision Fermi surface mapping reveals a complicated structure (Fig.~4a) which upon closer consideration is explained by crossed ellipses of the kind we observed in the case of electron-like pockets earlier (Fig.~4b). The presence of crossed ellipses in the center of the BZ has been also observed and discussed earlier~\cite{Watson2015a,Watson2015b}. The cut through the non-degenerate sections of these ellipses (red arrow in Fig.~4a) shown in Fig.~4c and its second derivative (Fig.~4d) unambiguously demonstrate that experimental electronic structure deviates from the calculated one. Taking into account that \textit{xy}-dispersion does not come close to the Fermi level (Fig.~1a), one clearly sees three other dispersing features in the immediate vicinity of the Fermi level, instead of two. It means that experimentally the $xz$/$yz$ dispersions are indeed different for two domains and this difference is strongly underestimated by the calculations. Since the results of our calculations are associated with the conventional phonon-driven structural transition, here we see the evidence for a much stronger effect which must be triggered by the interaction of different kind; in this case, obviously electronic in nature. 

Remarkably, the energy scale approximately coincides with the one extracted from the splittings at M-point: the energy separation of the dispersions which cross the Fermi level is of the order of 15 meV (see Fig.~4b). This is in line with the previous theoretical arguments of Fernandes and Vafek~\cite{FernandesVafek}. Indeed, the splitting of the bands in the center of the BZ and of bottom of the shallower electron pocket have been associated with the lifting of degeneracy of the $xz$/$yz$ bands, while the splitting of the bottom of the deeper electron pocket is due to the hopping anisotropy of $xy$ states. From Fig.~2g one can see that both are of the same order of magnitude with the latter being slightly smaller. Using the same approach~\cite{FernandesVafek} and knowing the energy scale of nematic order obtained in the present paper, we can determine the contribution of the spin-orbit interaction to the splitting of the $xz$/$yz$ bands in the center of the BZ more precisely. From the previously detected total splitting of 25 meV~\cite{Borisenko2015} only 20 meV are due to spin-orbit coupling ($25^{2}=20^{2}+15^{2}$). Exactly such splitting is directly observed in the center of the BZ above Ts.

\begin{figure}[h!]
	\centering
		\includegraphics[width=17cm]{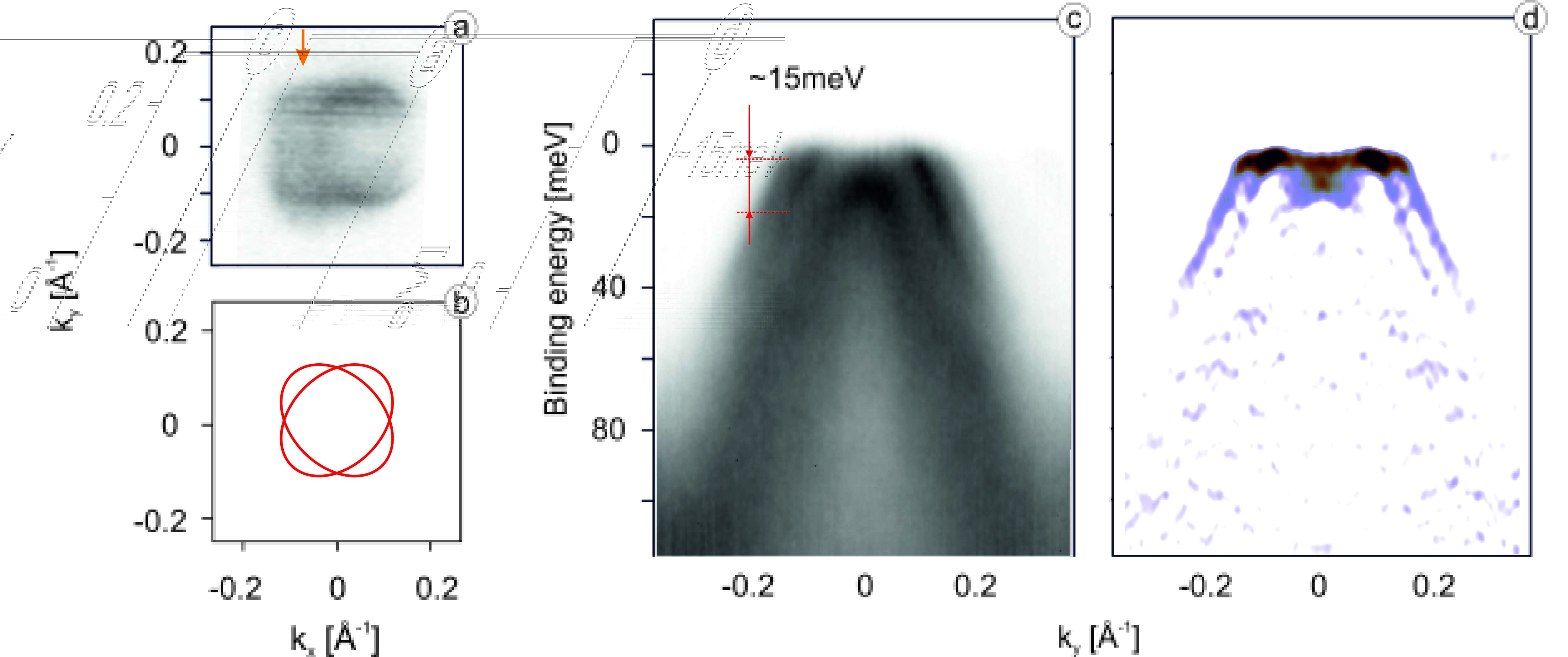}
	\caption{a) High resolution ARPES derived Fermi surface in vicinity of $Z$ point measured with $h\nu$=23eV; b) cartoon of the two domain Fermi surface in vicinity of $Z$ point; c) and d) high resolution energy cut along the shortest $Z-Z$ direction at k$_x$=0.05\AA$^{-1}$ and its second derivative respectively}
	\label{Fig:ElP}
\end{figure}

Experiment tells us that the modifications of the electronic structure of FeSe upon entering nematic phase are stronger near the center of the BZ than the simple lattice-distortion approach predicts, pointing to the electronic mechanism as a driving force. This implies a prevalence of the short momentum range interaction in formation of the nematic order in FeSe. In order to use this information to differentiate between the orbital or Ising-nematic orders as possible candidates, more rigorous quantitative estimates are needed.

\section{Conclusions}

In conclusion, our ARPES study demonstrated that the electronic structure of FeSe is indeed sensitive to the structural transition at 87K, however the energy scale has been overestimated earlier. The most of the features previously attributed to the nematic ordering in the vicinity of the corner of the BZ can be reproduced by the conventional band-structure calculations. Our calculations for the low temperature phase suggest more hidden effects of the band structure transformation which are supported by our ARPES results. In contrast, the experimentally derived electronic structure near the center of the BZ is not in agreement with simple DFT treatment. Namely, we found splitting of the hole pocket in the $\Gamma$ point originated by the presence of two domains in the orthorombic phase. Finally, we established the new energy and momentum scales of the nematic  order in FeSe which are consistent with the critical temperature of the transition, imply the electronic origin of the effect and provide a quantitative basis for differentiation between orbital and spin interactions as triggers of the nematic transition.

\textit{Note added:} While completing this work we became aware of a similar study on FeSe~\cite{Watson2016}. While the experimental data agree, and previously determined energy scale of 50 meV is reevaluated, interpretation of the details and conclusions are somewhat different.

\section{Experimental and computational details}

\subsection{Experimental details}
Samples were grown by the KCl/AlCl$_3$ chemical vapor transport method.
ARPES measurements were performed at the I05 beamline of Diamond Light Source, UK. Single-crystal samples were cleaved in situ in a vacuum lower
than 2$\times$10$^{-10}$ mbar and measured at temperatures ranging from 5.7 to 270 K. Measurements were performed using linearly
polarized synchrotron light with 23~eV and 28~eV for $\Gamma$ and $M$ point respectively, utilizing
Scienta R4000 hemispherical electron energy analyser with an angular resolution of 0.2$–$0.5 deg and an energy resolution of 3 meV.

\subsection{Computational details}

Band structure calculations were performed for the experimental crystal structures in the local density approximation (LSDA) using the linear muffin-tin orbital (LMTO) method. Spin-orbit coupling was added to the LMTO Hamiltonian at the variational step.

\section{Acknowledgements}

We are grateful to Andrey Chubukov, Matthew Watson, Rafael Fernandes, Brian Andersen, Amalia Coldea, Alexander Kordyuk, Laura Fanfarillo, Peter Hirschfeld, Roser Valenti, Sahana Roessler, Ilya Eremin and Seung-Ho Baek for the fruitful discussions. The work was supported under grants No. BO1912/2-2 and BO1912/3-1. A.F. and A.G. acknowledge the ERC grant no. 648589 'SUPER-2D'. A.F acknowledge the support of Saint Petersburg State University (research Grant No. 15.61.202.2015)

\bibliographystyle{naturemag}
\bibliography{paperbase}

\providecommand{\latin}[1]{#1}
\providecommand*\mcitethebibliography{\thebibliography}
\csname @ifundefined\endcsname{endmcitethebibliography}
  {\let\endmcitethebibliography\endthebibliography}{}
\begin{mcitethebibliography}{24}
\providecommand*\natexlab[1]{#1}
\providecommand*\mciteSetBstSublistMode[1]{}
\providecommand*\mciteSetBstMaxWidthForm[2]{}
\providecommand*\mciteBstWouldAddEndPuncttrue
  {\def\EndOfBibitem{\unskip.}}
\providecommand*\mciteBstWouldAddEndPunctfalse
  {\let\EndOfBibitem\relax}
\providecommand*\mciteSetBstMidEndSepPunct[3]{}
\providecommand*\mciteSetBstSublistLabelBeginEnd[3]{}
\providecommand*\EndOfBibitem{}
\mciteSetBstSublistMode{f}
\mciteSetBstMaxWidthForm{subitem}{(\alph{mcitesubitemcount})}
\mciteSetBstSublistLabelBeginEnd
  {\mcitemaxwidthsubitemform\space}
  {\relax}
  {\relax}

\bibitem[Fernandes \latin{et~al.}(2014)Fernandes, Chubukov, and
  Schmalian]{Fernandes2014}
Fernandes,~R.~M.; Chubukov,~A.~V.; Schmalian,~J. \emph{Nature Physics}
  \textbf{2014}, \emph{10}, 97--104\relax
\mciteBstWouldAddEndPuncttrue
\mciteSetBstMidEndSepPunct{\mcitedefaultmidpunct}
{\mcitedefaultendpunct}{\mcitedefaultseppunct}\relax
\EndOfBibitem
\bibitem[Fernandes and Millis(2013)Fernandes, and Millis]{Fernandes2013}
Fernandes,~R.~M.; Millis,~A.~J. \emph{Phys. Rev. Lett.} \textbf{2013},
  \emph{111}, 127001\relax
\mciteBstWouldAddEndPuncttrue
\mciteSetBstMidEndSepPunct{\mcitedefaultmidpunct}
{\mcitedefaultendpunct}{\mcitedefaultseppunct}\relax
\EndOfBibitem
\bibitem[Livanas \latin{et~al.}(2015)Livanas, Aperis, Kotetes, and
  Varelogiannis]{Livanas2015}
Livanas,~G.; Aperis,~A.; Kotetes,~P.; Varelogiannis,~G. \emph{Physical Review B
  - Condensed Matter and Materials Physics} \textbf{2015}, \emph{91},
  1--13\relax
\mciteBstWouldAddEndPuncttrue
\mciteSetBstMidEndSepPunct{\mcitedefaultmidpunct}
{\mcitedefaultendpunct}{\mcitedefaultseppunct}\relax
\EndOfBibitem
\bibitem[Yang \latin{et~al.}(2013)Yang, Wang, and Lee]{Yang2013}
Yang,~F.; Wang,~F.; Lee,~D.~H. \emph{Physical Review B - Condensed Matter and
  Materials Physics} \textbf{2013}, \emph{88}, 1--5\relax
\mciteBstWouldAddEndPuncttrue
\mciteSetBstMidEndSepPunct{\mcitedefaultmidpunct}
{\mcitedefaultendpunct}{\mcitedefaultseppunct}\relax
\EndOfBibitem
\bibitem[Baek \latin{et~al.}(2014)Baek, Efremov, Ok, Kim, van~den Brink, and
  B{\"{u}}chner]{Baek2014}
Baek,~S.-H.; Efremov,~D.~V.; Ok,~J.~M.; Kim,~J.~S.; van~den Brink,~J.;
  B{\"{u}}chner,~B. \emph{Nature Materials} \textbf{2014}, \emph{14},
  210--214\relax
\mciteBstWouldAddEndPuncttrue
\mciteSetBstMidEndSepPunct{\mcitedefaultmidpunct}
{\mcitedefaultendpunct}{\mcitedefaultseppunct}\relax
\EndOfBibitem
\bibitem[Zhang \latin{et~al.}(2015)Zhang, Qian, Richard, Wang, Miao, Lv, Fu,
  Wolf, Meingast, Wu, Wang, Hu, and Ding]{Zhang2015}
Zhang,~P.; Qian,~T.; Richard,~P.; Wang,~X.~P.; Miao,~H.; Lv,~B.~Q.; Fu,~B.~B.;
  Wolf,~T.; Meingast,~C.; Wu,~X.~X.; Wang,~Z.~Q.; Hu,~J.~P.; Ding,~H.
  \emph{Physical Review B - Condensed Matter and Materials Physics}
  \textbf{2015}, \emph{91}, 1--5\relax
\mciteBstWouldAddEndPuncttrue
\mciteSetBstMidEndSepPunct{\mcitedefaultmidpunct}
{\mcitedefaultendpunct}{\mcitedefaultseppunct}\relax
\EndOfBibitem
\bibitem[Maletz \latin{et~al.}(2014)Maletz, Zabolotnyy, Evtushinsky,
  Thirupathaiah, Wolter, Harnagea, Yaresko, Vasiliev, Chareev, B??hmer, Hardy,
  Wolf, Meingast, Rienks, B??chner, and Borisenko]{Maletz2014}
Maletz,~J. \latin{et~al.}  \emph{Physical Review B - Condensed Matter and
  Materials Physics} \textbf{2014}, \emph{89}, 1--5\relax
\mciteBstWouldAddEndPuncttrue
\mciteSetBstMidEndSepPunct{\mcitedefaultmidpunct}
{\mcitedefaultendpunct}{\mcitedefaultseppunct}\relax
\EndOfBibitem
\bibitem[Shimojima \latin{et~al.}(2014)Shimojima, Suzuki, Sonobe, Nakamura,
  Sakano, Omachi, Yoshioka, Kuwata-Gonokami, Ono, Kumigashira, B??hmer, Hardy,
  Wolf, Meingast, L??hneysen, Ikeda, and Ishizaka]{Shimojima2014}
Shimojima,~T. \latin{et~al.}  \emph{Physical Review B - Condensed Matter and
  Materials Physics} \textbf{2014}, \emph{90}, 1--5\relax
\mciteBstWouldAddEndPuncttrue
\mciteSetBstMidEndSepPunct{\mcitedefaultmidpunct}
{\mcitedefaultendpunct}{\mcitedefaultseppunct}\relax
\EndOfBibitem
\bibitem[Nakayama \latin{et~al.}(2014)Nakayama, Miyata, Phan, Sato, Tanabe,
  Urata, Tanigaki, and Takahashi]{Nakayama2014}
Nakayama,~K.; Miyata,~Y.; Phan,~G.~N.; Sato,~T.; Tanabe,~Y.; Urata,~T.;
  Tanigaki,~K.; Takahashi,~T. \emph{Physical Review Letters} \textbf{2014},
  \emph{113}, 1--5\relax
\mciteBstWouldAddEndPuncttrue
\mciteSetBstMidEndSepPunct{\mcitedefaultmidpunct}
{\mcitedefaultendpunct}{\mcitedefaultseppunct}\relax
\EndOfBibitem
\bibitem[Borisenko \latin{et~al.}(2016)Borisenko, Evtushinsky, Liu, Morozov,
  Kappenberger, Wurmehl, Buchner, Yaresko, Kim, Hoesch, Wolf, and
  Zhigadlo]{Borisenko2015}
Borisenko,~S.~V.; Evtushinsky,~D.~V.; Liu,~Z.-H.; Morozov,~I.;
  Kappenberger,~R.; Wurmehl,~S.; Buchner,~B.; Yaresko,~A.~N.; Kim,~T.~K.;
  Hoesch,~M.; Wolf,~T.; Zhigadlo,~N.~D. \emph{Nat Phys} \textbf{2016},
  \emph{12}, 311--317\relax
\mciteBstWouldAddEndPuncttrue
\mciteSetBstMidEndSepPunct{\mcitedefaultmidpunct}
{\mcitedefaultendpunct}{\mcitedefaultseppunct}\relax
\EndOfBibitem
\bibitem[Watson \latin{et~al.}(2015)Watson, Kim, Haghighirad, Davies, McCollam,
  Narayanan, Blake, Chen, Ghannadzadeh, Schofield, Hoesch, Meingast, Wolf, and
  Coldea]{Watson2015a}
Watson,~M.~D.; Kim,~T.~K.; Haghighirad,~A.~A.; Davies,~N.~R.; McCollam,~A.;
  Narayanan,~A.; Blake,~S.~F.; Chen,~Y.~L.; Ghannadzadeh,~S.; Schofield,~A.~J.;
  Hoesch,~M.; Meingast,~C.; Wolf,~T.; Coldea,~A.~I. \emph{Physical Review B -
  Condensed Matter and Materials Physics} \textbf{2015}, \emph{91}, 1--14\relax
\mciteBstWouldAddEndPuncttrue
\mciteSetBstMidEndSepPunct{\mcitedefaultmidpunct}
{\mcitedefaultendpunct}{\mcitedefaultseppunct}\relax
\EndOfBibitem
\bibitem[Fanfarillo \latin{et~al.}(2016)Fanfarillo, Mansart, Toulemonde,
  Cercellier, Fevre, Bertran, Valenzuela, Benfatto, and Brouet]{Fanfarillo2016}
Fanfarillo,~L.; Mansart,~J.; Toulemonde,~P.; Cercellier,~H.; Fevre,~P.~L.;
  Bertran,~F.; Valenzuela,~B.; Benfatto,~L.; Brouet,~V. \emph{Arxiv}
  \textbf{2016}, 1--16\relax
\mciteBstWouldAddEndPuncttrue
\mciteSetBstMidEndSepPunct{\mcitedefaultmidpunct}
{\mcitedefaultendpunct}{\mcitedefaultseppunct}\relax
\EndOfBibitem
\bibitem[Xu \latin{et~al.}(2016)Xu, Niu, Xu, Jiang, Yao, Abdel-Hafiez, Chareev,
  Vasiliev, Peng, and Feng]{Xu2016}
Xu,~H.~C.; Niu,~X.~H.; Xu,~D.~F.; Jiang,~J.; Yao,~Q.; Abdel-Hafiez,~M.;
  Chareev,~D.~A.; Vasiliev,~A.~N.; Peng,~R.; Feng,~D.~L. \emph{arXiv}
  \textbf{2016}, \emph{1}, 1--5\relax
\mciteBstWouldAddEndPuncttrue
\mciteSetBstMidEndSepPunct{\mcitedefaultmidpunct}
{\mcitedefaultendpunct}{\mcitedefaultseppunct}\relax
\EndOfBibitem
\bibitem[Ye \latin{et~al.}(2015)Ye, Zhang, Ning, Li, Chen, Jia, Hashimoto, Lu,
  Shen, and Zhang]{Ye2015}
Ye,~Z.~R.; Zhang,~C.~F.; Ning,~H.~L.; Li,~W.; Chen,~L.; Jia,~T.; Hashimoto,~M.;
  Lu,~D.~H.; Shen,~Z.~X.; Zhang,~Y. \emph{Arxiv Preprint} \textbf{2015},
  1512.02526\relax
\mciteBstWouldAddEndPuncttrue
\mciteSetBstMidEndSepPunct{\mcitedefaultmidpunct}
{\mcitedefaultendpunct}{\mcitedefaultseppunct}\relax
\EndOfBibitem
\bibitem[Kotliar \latin{et~al.}(2006)Kotliar, Savrasov, Haule, Oudovenko,
  Parcollet, and Marianetti]{Kotliar2006}
Kotliar,~G.; Savrasov,~S.~Y.; Haule,~K.; Oudovenko,~V.~S.; Parcollet,~O.;
  Marianetti,~C.~A. \emph{Reviews of Modern Physics} \textbf{2006}, \emph{78},
  865--951\relax
\mciteBstWouldAddEndPuncttrue
\mciteSetBstMidEndSepPunct{\mcitedefaultmidpunct}
{\mcitedefaultendpunct}{\mcitedefaultseppunct}\relax
\EndOfBibitem
\bibitem[Ferber \latin{et~al.}(2012)Ferber, Foyevtsova, Valent\'{\i}, and
  Jeschke]{Valenti2012}
Ferber,~J.; Foyevtsova,~K.; Valent\'{\i},~R.; Jeschke,~H.~O. \emph{Phys. Rev.
  B} \textbf{2012}, \emph{85}, 094505\relax
\mciteBstWouldAddEndPuncttrue
\mciteSetBstMidEndSepPunct{\mcitedefaultmidpunct}
{\mcitedefaultendpunct}{\mcitedefaultseppunct}\relax
\EndOfBibitem
\bibitem[Chubukov \latin{et~al.}(2016)Chubukov, Khodas, and
  Fernandes]{Chubukov2016}
Chubukov,~A.~V.; Khodas,~M.; Fernandes,~R.~M. \emph{arXiv} \textbf{2016},
  \emph{1}, 1--61\relax
\mciteBstWouldAddEndPuncttrue
\mciteSetBstMidEndSepPunct{\mcitedefaultmidpunct}
{\mcitedefaultendpunct}{\mcitedefaultseppunct}\relax
\EndOfBibitem
\bibitem[Ortenzi \latin{et~al.}(2009)Ortenzi, Cappelluti, Benfatto, and
  Pietronero]{BenfattoPhysRevLett.103.046404}
Ortenzi,~L.; Cappelluti,~E.; Benfatto,~L.; Pietronero,~L. \emph{Phys. Rev.
  Lett.} \textbf{2009}, \emph{103}, 046404\relax
\mciteBstWouldAddEndPuncttrue
\mciteSetBstMidEndSepPunct{\mcitedefaultmidpunct}
{\mcitedefaultendpunct}{\mcitedefaultseppunct}\relax
\EndOfBibitem
\bibitem[Charnukha \latin{et~al.}(2015)Charnukha, Evtushinsky, Matt, Xu, Shi,
  B{\"{u}}chner, Zhigadlo, Batlogg, and Borisenko]{Charnukha2015}
Charnukha,~A.; Evtushinsky,~D.~V.; Matt,~C.~E.; Xu,~N.; Shi,~M.;
  B{\"{u}}chner,~B.; Zhigadlo,~N.~D.; Batlogg,~B.; Borisenko,~S.~V.
  \emph{Scientific Reports} \textbf{2015}, \emph{5}, 18273\relax
\mciteBstWouldAddEndPuncttrue
\mciteSetBstMidEndSepPunct{\mcitedefaultmidpunct}
{\mcitedefaultendpunct}{\mcitedefaultseppunct}\relax
\EndOfBibitem
\bibitem[Khasanov \latin{et~al.}(2010)Khasanov, Bandele, Conder, Keller,
  Pomjakushina, and Pomjakushin]{Khasanov2010}
Khasanov,~R.; Bandele,~M.; Conder,~K.; Keller,~H.; Pomjakushina,~E.;
  Pomjakushin,~V. \emph{New Journal of Physics} \textbf{2010}, \emph{12}\relax
\mciteBstWouldAddEndPuncttrue
\mciteSetBstMidEndSepPunct{\mcitedefaultmidpunct}
{\mcitedefaultendpunct}{\mcitedefaultseppunct}\relax
\EndOfBibitem
\bibitem[Watson \latin{et~al.}(2015)Watson, Kim, Haghighirad, Blake, Davies,
  Hoesch, Wolf, and Coldea]{Watson2015b}
Watson,~M.~D.; Kim,~T.~K.; Haghighirad,~A.~A.; Blake,~S.~F.; Davies,~N.~R.;
  Hoesch,~M.; Wolf,~T.; Coldea,~A.~I. \emph{Phys. Rev. B} \textbf{2015},
  \emph{92}, 121108\relax
\mciteBstWouldAddEndPuncttrue
\mciteSetBstMidEndSepPunct{\mcitedefaultmidpunct}
{\mcitedefaultendpunct}{\mcitedefaultseppunct}\relax
\EndOfBibitem
\bibitem[Fernandes and Vafek(2014)Fernandes, and Vafek]{FernandesVafek}
Fernandes,~R.~M.; Vafek,~O. \emph{Phys. Rev. B} \textbf{2014}, \emph{90},
  214514\relax
\mciteBstWouldAddEndPuncttrue
\mciteSetBstMidEndSepPunct{\mcitedefaultmidpunct}
{\mcitedefaultendpunct}{\mcitedefaultseppunct}\relax
\EndOfBibitem
\bibitem[Watson \latin{et~al.}(2016)Watson, Kim, Hoesch, Haghighirad, and
  Coldea]{Watson2016}
Watson,~M.~D.; Kim,~T.~K.; Hoesch,~M.; Haghighirad,~A.~A.; Coldea,~A.~I.
  \emph{arXiv} \textbf{2016}, 1603.04545\relax
\mciteBstWouldAddEndPuncttrue
\mciteSetBstMidEndSepPunct{\mcitedefaultmidpunct}
{\mcitedefaultendpunct}{\mcitedefaultseppunct}\relax
\EndOfBibitem
\end{mcitethebibliography}
\end{document}